\documentclass[9pt,conference]{IEEEtran}
\usepackage{dcase2026}

\usepackage{bm} 

\usepackage{dcase2026,amsmath,graphicx,url,times,booktabs, tabularx}

\title{STFT Frontend Mismatch in Pretrained Audio Spectrogram Transformers: Analysis and Adaptation}

\name{Julia Wilkins$^{1}$,
      Wei-Ching Lin$^{2}$,
      Abinaya Kumar$^{2}$,
      Shabnam Ghaffarzadegan$^{2}$,
      Juan Pablo Bello$^{1}$}
\address{
    $^1$ New York University, USA \\
    $^2$ Bosch Research, USA\\
}

\begin{document}

\maketitle

\begin{abstract}


Pretrained audio spectrogram transformers are widely used as backbones of machine listening systems for sound event classification. However, they are typically trained using spectrograms with a fixed Short-Time Fourier Transform (STFT) configuration, implicitly binding the model to the time-frequency resolution seen during pretraining. In practice, adapting these systems to new domains or computational constraints often requires different STFT frontend parameters. In this work, we conduct a controlled study of the sensitivity of the foundational Audio Spectrogram Transformer to variations in STFT window and hop size, showing that audio classification performance degrades substantially as frontend parameters deviate from pretraining settings. As a mitigation strategy, we introduce a lightweight conditioning mechanism using FiLM that enables a pretrained model to adapt to variable input configurations using limited additional training data. Our method generalizes well to unseen STFT configurations, improving model robustness without sacrificing performance in the original setting.




\end{abstract}


\section{Introduction}

Pretrained transformer-based models have become foundational to many machine listening systems, with the Audio Spectrogram Transformer (AST) \cite{gong2021ast} and its variants \cite{gong2022ssast, koutini2022patchout, chen2022hts, baade2022mae} emerging as widely adopted backbones in the audio domain due to strong performance on audio classification benchmarks \cite{piczak2015esc, gemmeke2017audio} and ability to capture long-range temporal structure \cite{vaswani2017attention}. AST operates on log-Mel spectrograms as input, computed via the Short-Time Fourier Transform (STFT) \cite{dennis2010spectrogram, muller2015fundamentals}. 
STFT parameters, such as window and hop length, control the time-frequency resolution of the resulting spectrogram for a given sampling rate \cite{muller2015fundamentals}. This comes with fundamental tradeoffs: longer windows improve frequency resolution but reduce temporal precision and vice-versa; shorter hops improve temporal precision but increase computational costs. AST, like most pretrained machine listening models, is trained with a fixed STFT configuration, implicitly tying the model to the time-frequency resolution seen during pretraining.

In practice, when pretrained AST models are applied to new downstream tasks and datasets, the STFT hyperparameters are typically inherited from the pretraining pipeline rather than treated as a deliberate design decision. Yet audio datasets vary considerably in their acoustic characteristics: some tasks demand fine temporal resolution to capture transient sounds or rapid transitions \cite{khunarsal2013very, huzaifah2017comparison, Liu_Liu_Kong_Wang_Plumbley_2024}, such as phoneme boundaries \cite{toledano2018multi, zhu2024wavrx} or species call onsets \cite{ ulloa2016screening, lostanlen2019long, knight2020pre}, while computational constraints in low-latency or resource-limited settings may necessitate coarser representations \cite{Martinmorato2022lowcomplexity, yeow2024real}. When these demands conflict with the pretrained frontend configuration, the result is what we refer to in this study as an \textit{STFT frontend mismatch}: a systematic discrepancy between the time-frequency resolution assumed by the model and what the downstream dataset or task requires.

This mismatch is typically treated as an input shape incompatibility and addressed with superficial workarounds. Common approaches include cropping, padding, repeating, or interpolating the input spectrogram to match the pretrained model's expected dimensions, which can distort the time-frequency structure or discard information \cite{chen2022hts, kong2020panns}. Alternatively, one could fine-tune or fully retrain the model on a new STFT configuration, but beyond the substantial data and compute requirements (e.g., AST was trained on 2 million 10s clips for 600 hours on 4 GPUs \cite{gong2021ast, chen2022hts}), this simply shifts the constraint to a different fixed configuration and does not ensure robustness to other unseen resolutions.

Prior work has been done to improve the flexibility of AST to changes in spectrogram patch size \cite{feng2023flexiast} and audio input length or temporal resolution \cite{feng2024elasticast}. The former uses variable patch sizes to proportionally adjust patch embedding weights during training \cite{feng2023flexiast}. The latter introduces a sequence packing method with masked self-attention to allow AST to handle variable-length inputs and temporally compressed resolutions \cite{feng2024elasticast}. While promising, these approaches still require re-training or fine-tuning a significant portion of the original AST due to underlying architecture changes. Further, neither approach directly investigates AST's inflexibility to changes in both window and hop length simultaneously. 
Other similarly motivated works propose a module for learning the optimal temporal resolution of a spectrogram for classification tasks \cite{Liu_Liu_Kong_Wang_Plumbley_2024}, or designing CNN layers that generalize across sampling rates for source separation \cite{saito2022sampling}; the prior focuses on resolution selection during model training rather than pretrained model adaptation and the latter on sampling rate invariance alone. We address these limitations via the following contributions in this work:

\begin{itemize}

    \item We present a systematic analysis of the sensitivity of pretrained AST to STFT frontend mismatch within a controlled study, showing substantial audio classification performance degradation as window and hop length deviate from pretraining settings.
   
    \item We propose a lightweight FiLM-based conditioning mechanism that enables pretrained AST to adapt to variable STFT window and hop lengths with minimal additional training.

     \item We show that our adaptation approach improves audio classification performance across unseen frontend settings while maintaining strong performance in the original configuration.
     
    \item We conduct a data efficiency study and show that substantial performance improvements can be achieved with as few as $1,000$ additional training samples with our method.

\end{itemize}

For the scope of this work, we analyze the original AST model \cite{gong2021ast} as a widely adopted backbone for audio classification tasks. We use AudioSet \cite{gemmeke2017audio} as an in-domain dataset to limit additional domain shift effects. We use STFT window and hop length as the axes of frontend variation, controlling for the sampling rate and other STFT hyperparameters to reduce confounding factors.

\section{Characterizing STFT Frontend Sensitivity}

\begin{figure}
    \centering
    \includegraphics[width=\linewidth,trim={1.5cm 0cm 0.5cm 0cm}]{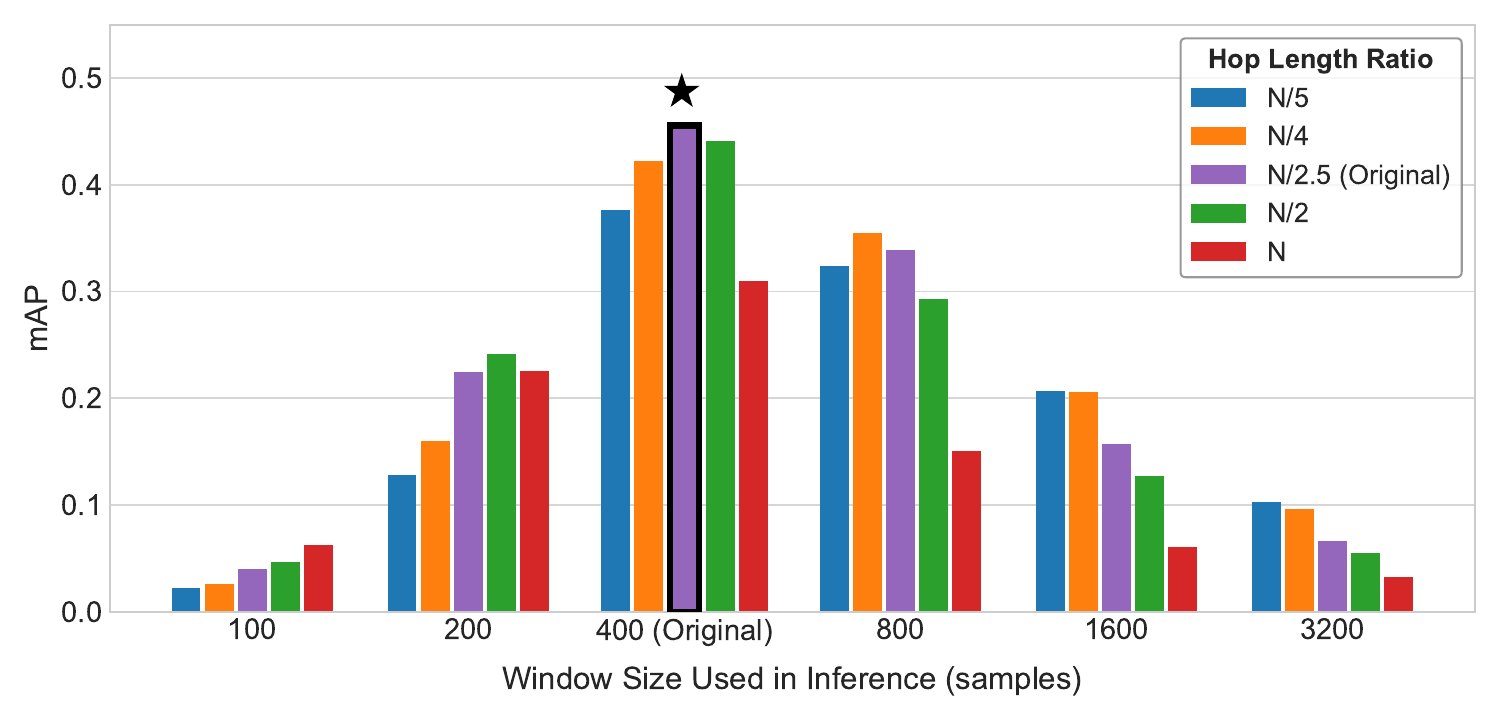}
    \caption{Audio classification performance using AST degrades significantly as window and hop length parameters deviate from the original configurations used in pretraining (denoted with $\mathbf{\star}$).}
    \label{fig:motivation}
\end{figure}

To quantify AST's sensitivity to STFT frontend mismatch, we first perform zero-shot evaluation using a pretrained Audio Spectrogram Transformer \cite{gong2021ast} (AST) across a range of STFT window and hop lengths. For each configuration, we compute audio classification performance on the AudioSet \cite{gemmeke2017audio} evaluation set, and measure this relative to performance under the settings used during pretraining.
The findings in Figure \ref{fig:motivation} illustrate our core motivating observation: AST performance degrades significantly when the window and hop length of the STFT used to create the input spectrogram differ from those used during pretraining. The original pretraining STFT configuration ($\star$) achieves the highest mean average precision (mAP) at $0.456$, and classification performance consistently declines as window and hop length diverge from this matched setting. Reducing the window size by only $12.5$ ms while maintaining the original window-hop length ratio yields over a $50\%$ relative decrease in mAP  ($0.456{\rightarrow}0.224$). Under more extreme shifts in time-frequency resolution such as with a window size of $100$ samples, or a hop size equal to the window size ($N$), performance often falls below $0.1$.

This analysis provides a systematic characterization of AST's sensitivity to frontend mismatch, an effect that has not been explicitly quantified previously in terms of STFT window and hop length variation in pretrained AST models. This also establishes the zero-shot degradation baselines against which we will measure our adaptation method, and directly motivates the need for an adaptation strategy that allows pretrained AST models to operate more robustly across varying frontend configurations without large-scale re-training.

\vspace{-0.1cm}

\section{STFT Frontend Adaptation}
To mitigate the degradation identified in Section 2, we propose a lightweight framework for variable frontend adaptation in pretrained AST models using FiLM conditioning \cite{perez2018film}. FiLM is well suited to this problem because it enables conditional modulation of intermediate model representations using only a small number of additional learnable parameters, and has proven successful in contexts such as multi-modal source separation \cite{liu2022separate, liu2024separate}, speech recognition \cite{yang2022film}, and speech robustness to temporal changes \cite{wisnu2025stsm}. We build on the original AST \cite{gong2021ast} as a simple, interpretable backbone for studying STFT frontend adaptation in isolation, without additional architectural or data augmentation changes.

\subsection{Audio Spectrogram Transformer}

AST \cite{gong2021ast} takes a log-Mel spectrogram $x \in \mathbb{R}^{f \times t}$ of an audio clip as input, where $f$ is the number of Mel-frequency bins and $t$ is the number of temporal frames determined by the STFT window and hop size. The spectrogram is split into a sequence of overlapping patches, which are then flattened to a series of $1$D patch embeddings via a linear projection. We denote the patch embeddings as $E \in \mathbb{R}^{T \times 768}$, where $T$ is the number of patches and $768$ is the embedding dimension. \texttt{[CLS]} and \texttt{[DIST]} tokens are appended to the beginning of the patch embedding sequence, and a learnable positional encoding $P\in \mathbb{R}^{(T+2) \times 768}$ is added to each patch embedding. This sequence is the input to a $12$-layer Transformer encoder \cite{vaswani2017attention}. The \texttt{[CLS]} and \texttt{[DIST]} outputs of the transformer are averaged and fed through a linear classifier for the supervised audio classification task. AST is originally trained on AudioSet-unbalanced \cite{gemmeke2017audio} and uses pretrained weights adapted from ViT \cite{dosovitskiy2020image}.

\subsection{Multi-scale Spectrogram Configuration Training}
\label{subsec:multiscalemethod}
To isolate the effect of changes to the time-frequency resolution of the input log-Mel spectrogram, we vary only the STFT window length ($N$) and hop length ($h$) in our study. We keep all other spectrogram parameters fixed, namely sample rate (\SI{16}{\kilo\hertz}), number of Mel-frequency bins ($128$), and window type (Hann), and leave variation in these parameters for future studies.

During large-scale pretraining, AST uses a window length of $N{=} 400$ and hop length of $h{=}160$, in samples at  \SI{16}{\kilo\hertz}. This corresponds to $25$ms and $10$ms, respectively, with a window-hop ratio of $N/2.5$. We probe window lengths logarithmically spaced around (but excluding) the reference of $400$ samples, and a set of hop lengths proportional to window length. Specifically, we define the set of possible window lengths $\mathcal{N} \in \{100, 200, 800, 1600, 3200\}$ samples ($\{6.25, 12.5, 50, 100, 200\}$ms, equivalently). For each window length $N \in \mathcal {N}$, we consider a set of hop lengths as a function of $N$: $\mathcal{H}(N) \in \{N/5, N/4, N/2, N\}$, varying the window-hop ratio while preserving relative scale across window lengths. We do not include the original window and hop length parameters ($N=400, h=N/2.5$) in the set of training configurations, as we aim to train and evaluate using fully unseen parameters.


We create a 5-fold cross-validation protocol over the set of window lengths $\mathcal{N}$. In each fold, three window lengths are used as options for training, one for validation, and one for testing. All subsets are disjoint.
For example, one fold uses window lengths $\{100, 1600, 3200\}$ for training, $\{800\}$ for validation, and $\{200\}$ for test. Across the five folds, each window length in $\mathcal{N}$ is used exactly once as the held-out test configuration, ensuring that performance is evaluated on all window lengths under unseen conditions. During training, for each mini-batch, we randomly sample a window length $N$ from the training subset of $\mathcal{N}$, and then randomly sample hop length $h \in \mathcal{H}(N)$. The selected configuration $\mathcal{C} = \{N, h\}$ is used to parameterize the STFT for that batch, thus exposing the model to multiple time-frequency resolutions throughout training. This paradigm is loosely akin to the Multi-Scale Spectral Loss \cite{engel2020ddsp, schwar2023multi}, but serves a  different purpose, as a mechanism used to generate variable input representations and corresponding conditioning signals vs. a reconstruction-driven objective.


\subsection{Conditioning with FiLM}

To adapt pretrained, frozen AST to unseen spectrogram configurations using minimal additional training parameters and limited training data, we propose to modulate the pretrained AST model using Feature-wise Linear Modulation (FiLM) \cite{perez2018film}. FiLM applies feature-wise affine transformations to intermediate layers of a network given conditioning input. With FiLM, conditioning inputs are mapped to scaling and shifting coefficients via a learnable sub-network. These coefficients can then be directly applied as transformations to modulate the features of another network, taking the standard form: $\text{FiLM}(\textbf{x}) = \gamma(\textbf{z}) \odot \textbf{x} + \beta(\textbf{z})$,
where $\textbf{x}$ is the input to the FiLM layer and $\gamma$ and $\beta$ are scaling and shifting vectors that depend on the conditioning input $\textbf{z}$.

For our application, the conditioning information is the window and hop length, chosen as the configuration for a particular training batch and parameterized as $\textbf{z} = [\text{log}(N), \text{log}(h)]$. We observed that using the logarithm of the window and hop size vs. the raw values slightly improved training stability and performance, likely because of the multiplicative nature of changes in time-frequency resolution. 

We freeze the entire AST model (${\approx}87$ million parameters) and train only the additional FiLM layers ($198,528$ parameters) for our experiments. The FiLM sub-network contains two linear layers with a hidden dimension of $128$ and GeLU activation \cite{hendrycks2016gaussian}, projecting $\mathbf{z}$ to dimension $1536$ such that it can be split in half to form scaling and shifting parameters $\gamma$ and $\beta$. We experiment with applying FiLM along the embedding dimension at three different points in the pretrained AST model ($\textbf{x}$ in the FiLM equation above): (1) to the patch embeddings (FiLM-PatchEmb), (2) the positional encoding (FiLM-PosEnc), and (3) within each layer of the Transformer encoder (FiLM-Encoder). Notably, in each of these settings the FiLM network contains the same number of parameters due to its feature-wise nature. 

When we use FiLM inside the Transformer encoder layers, we learn a single set of parameters and share that across each layer of the encoder. Additionally, in this setting we experiment with different FiLM locations \textit{within} the Transformer layers: after initial layer normalization, after multi-head attention, after layer normalization prior to the feed-forward layer, and after the final feed-forward layer. We find applying FiLM prior to the feed-forward layer to be  most effective and report results for this setting as our best FiLM-Encoder in Section 5.

\vspace{-0.15cm}
\section{Experimental Design}
\subsection{Dataset}
We choose AudioSet \cite{gemmeke2017audio} for this study because it was the pretraining dataset for AST \cite{gong2021ast}, allowing us to isolate the effect of variable STFT frontend configurations without confounding domain shift effects. This also makes zero-shot evaluation reasonable. We use the balanced subset of AudioSet for all model training, to illustrate model adaptation with minimal additional training resources. This subset contains $19,644$ audio clips multi-labeled across $527$ sound event classes. For evaluation, we use the AudioSet evaluation set containing $17,979$ clips\footnote{The original balanced and evaluation AudioSet contain $22$k and ~$20$k clips, respectively. Due to changes in privacy restrictions and removal of some videos, at the time of this study we were not able to obtain the entirety of the dataset.}. Each audio clip is $10$ seconds long and sampled at \SI{16}{\kilo\hertz}, following the original AST.
We randomly split the AudioSet balanced (train) dataset into training and validation subsets, containing $16,697$ and $2,947$ clips respectively ($85{:}15$ split). We use the full AudioSet evaluation set for testing throughout. 

\subsection{Model Adaptation Considerations}
Because AST is trained on $10$-second clips with a fixed STFT configuration, the learned positional encoding is tied to the resulting token length of the original patched spectrogram. Modifying the window or hop length changes the number of temporal frames and thus the number of patch embeddings, causing a shape mismatch when adding the pretrained positional encoding to the patch embeddings. In the original AST preprocessing, such mismatches are handled by zero-padding or cropping the input spectrogram. However, in our case this could lead to information loss especially when smaller hop sizes produce denser temporal representations (e.g., more frames). Instead, we interpolate or crop the pretrained positional encoding to match the incoming patch embedding sequence, following the adaptation strategy used in \cite{gong2021ast} to transfer ViT \cite{dosovitskiy2020image} positional encodings to AST. This is the only modified component of the original AST architecture in our proposed method, and is an inference-time scaling operation and not a trainable component.

\subsection{Training Details}
We follow the $5$-fold multi-scale spectrogram cross-validation setup described in Section \ref{subsec:multiscalemethod}. The pretrained base model used throughout is the base AST (patch size $16$). Each model is trained for $30$ epochs with a batch size of $8$, which takes $3{-}4$ hours on $1\times$V100 GPU. For the few-shot adaptation experiments, we train for a maximum of $100$ epochs using a random subset (of the full balanced set of $\sim20$k) of $10$, $100$, $1$k, $2$k, $5$k, and $10$k samples.
For all experiments, we use the Adam optimizer \cite{kingma2014adam} with an initial learning rate of $2.5e{-}4$ and exponential decay. 
We do not use additional data augmentations as in the original AST in order to best isolate our adaptation effect. The training is supervised using a binary cross-entropy objective. Model selection is performed using lowest validation loss during training.

\subsection{Evaluation Metrics}
Each of the $5$ folds as defined in \ref{subsec:multiscalemethod} has exactly one held-out window size. For each model architecture and fold, we evaluate using the held-out window size and $4$ window-hop ratios associated with this window size for inference. This yields $4$ results per fold, totaling to $|\mathcal{C}|=4{\times} 5 =20$ window-hop configurations to evaluate per model architecture. We evaluate each of the $20$ configurations above in terms of mean average precision (mAP) per the original multi-label evaluation protocol for classification with AudioSet \cite{gemmeke2017audio} used in AST.

\subsection{Baseline Methods}

We use the pretrained AST model, frozen without additional training, as the \textbf{zero-shot} baseline when evaluated on STFT window and hop configurations \textit{outside} the pretraining setting. This quantifies performance degradation introduced by frontend mismatch and serves as the primary reference point for measuring the benefits of our adaptation method across new configurations (Figure {\ref{fig:motivation}}). The zero-shot baseline, evaluated using the \textit{original} STFT window and hop length ($\mathcal{C}=\{400,160\}$) used in pretraining, serves as the \textbf{upper-bound}. This functions as a topline, reflecting performance without frontend mismatch.

As a lightweight adaptation alternative to FiLM, we provide \textbf{fine-tuning} (FT) baselines, in which either the classifier head (FT-ClfHead) or the positional encoding (FT-PosEnc) is fine-tuned and the rest of the pretrained AST is kept frozen. The fine-tuning is done using the identical multi-scale spectrogram training regime as is used for our proposed FiLM adapter. We do not include a fully fine-tuned or trained from scratch version of AST in the scope of this study, as we aim to provide computationally efficient adaptation strategies; we also suspect that training in this setting given the size of AST relative to the small size of AudioSet-balanced would not be particularly effective.

\begin{figure*}[t]
    \label{fig:robustness_and_preservation}

    \centering

    \begin{minipage}{0.66\linewidth}
        \centering
        \includegraphics[height=5.9cm, trim={1.5cm 0 0 0.5cm},width=\linewidth]{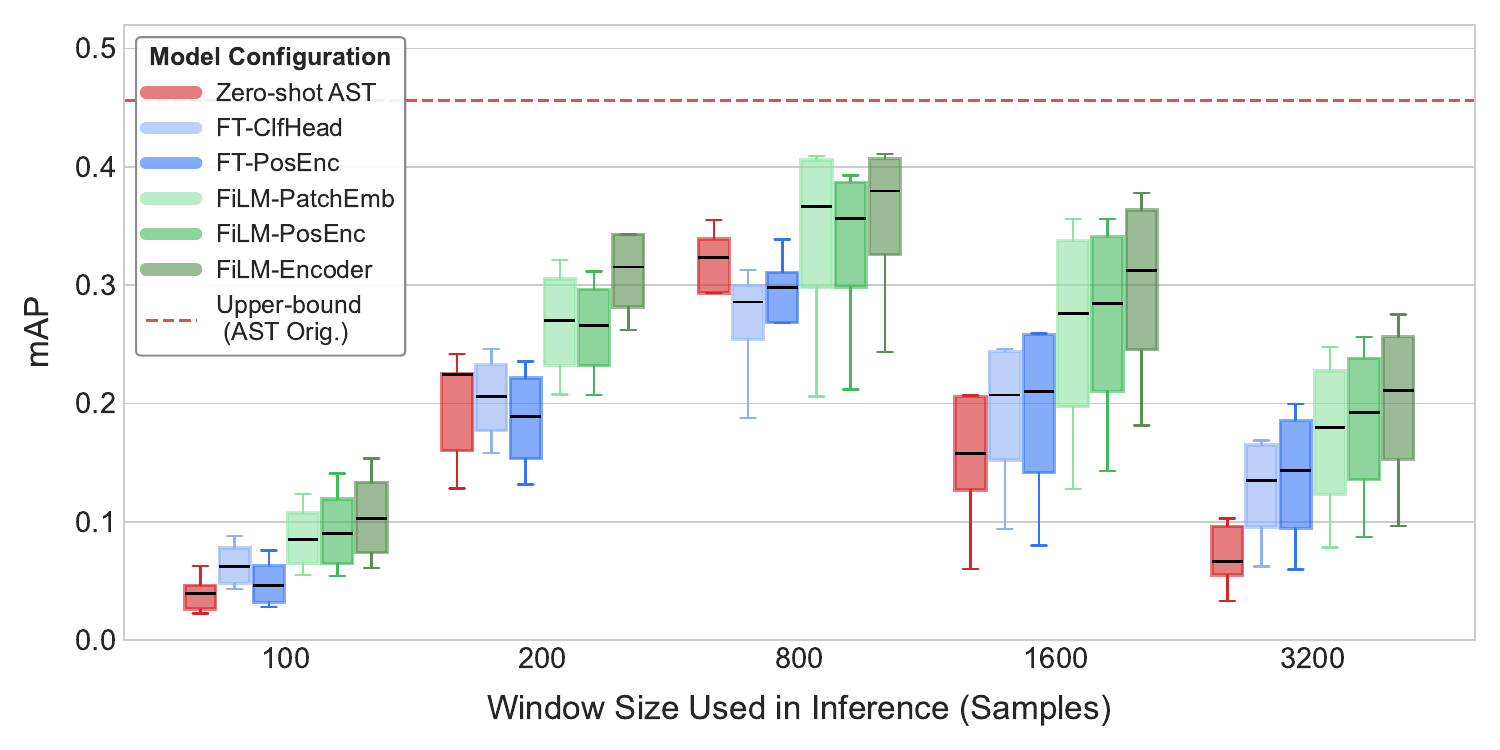}
        \vspace{-0.5cm}
    \end{minipage}
    \hfill
    \begin{minipage}{0.32\linewidth}
        \centering
        \includegraphics[trim={0 0 0 1.5cm},width=\linewidth]{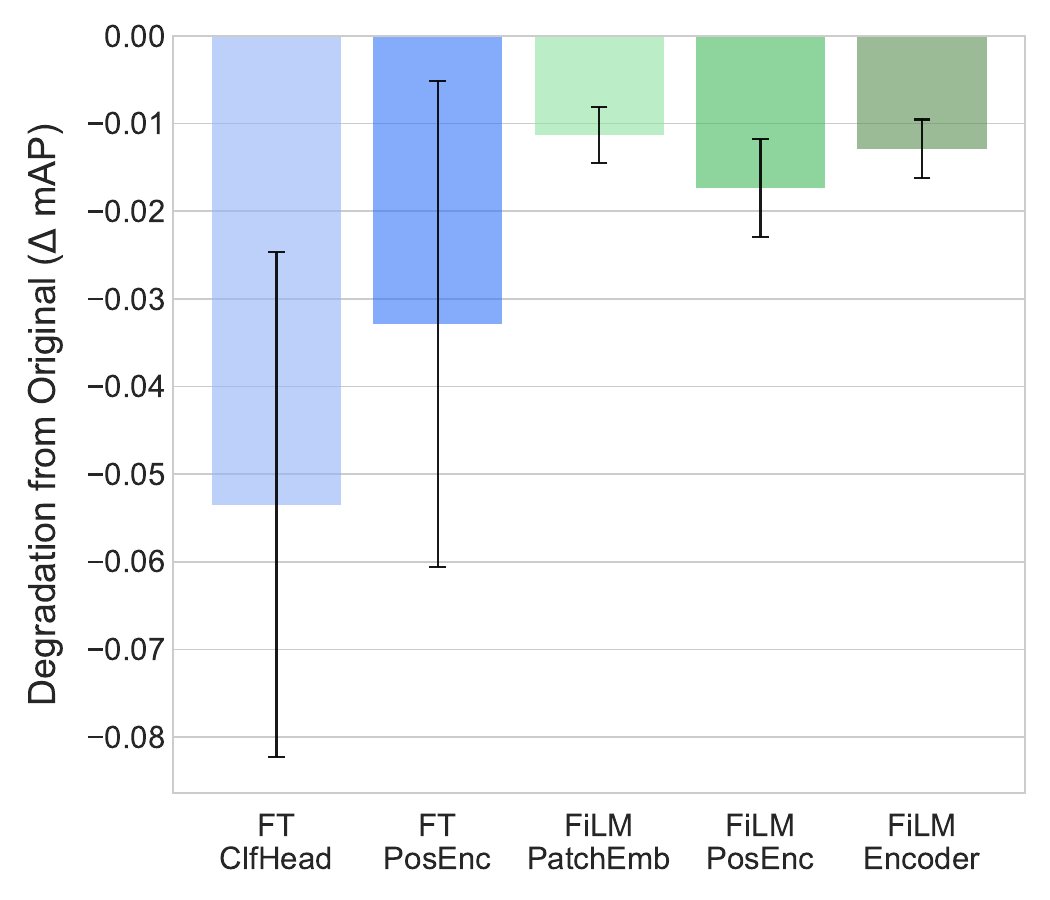}
        \vspace{-0.3cm}
    \end{minipage}

    \caption{
    (\textbf{Left}): Performance on unseen spectrogram configurations, in terms of mAP for each window length across window-hop ratios. 
    (\textbf{Right}): Performance degradation relative to the original AST upper-bound ($\text{mAP}-0.456$), shown across folds per model configuration.
    }
\end{figure*}

\section{Results}

\subsection{Adapting AST to unseen configurations with FiLM}

Figure \ref{fig:robustness_and_preservation} (Left) shows audio classification performance (mAP) for our proposed FiLM variants compared to baseline methods, as a function of inference-time window length across window-hop ratios. The zero-shot AST baseline (red) shows a significant decline in performance from the matched upper-bound as window length deviates from the pretrained configuration ($N{=}400$), particularly at more extreme temporal resolutions (e.g., $N{=}100,3200$). Fine-tuning baselines (blue) occasionally offer a slight improvement over zero-shot, especially when using larger window lengths. However, when the inference window length is closest to the original, fine-tuning slightly hurts performance relative to zero-shot. This suggests a tradeoff; fine-tuning using multiple frontend configurations may shift the learned distribution away from the pretrained optimum, reducing performance in matched or near-matched settings. We do not observe a significant difference between the effects of fine-tuning the classification head vs. the positional encoding.

In contrast, our FiLM-conditioned models (green) demonstrate consistent improvement across all evaluated unseen window lengths.
The FiLM variants outperform fine-tuning baselines throughout and substantially improve upon the degradation observed in the zero-shot setting. We find that applying FiLM within the Transformer encoder layers (specifically prior to each feed-forward layer) is the most effective strategy. At $N{=}1600$, this configuration (FiLM-Encoder) nearly doubles zero-shot performance ($0.166 {\rightarrow} 0.313$, $+14.7$pp) in terms of median mAP across hop lengths per window. For window lengths closer to the pretrained setting ($N{=}200, 800$), FiLM still yields substantial gains, improving over zero-shot by $+12.3$ and $+7.1$pp, respectively. Notably, these improvements occur in both interpolated and extrapolated settings (e.g., within-range and out-of-range of window lengths seen during adaptation training), demonstrating that FiLM conditioning offers a strong, generalizable approach to enhancing robustness to frontend mismatch in pretrained ASTs. 

In general, we also observe a large spread in results across hop ratios per window length in Figure \ref{fig:robustness_and_preservation} (Left). The smallest ($N/5$) and largest ($N$) hop settings consistently exhibit the lowest performance, likely due to the scale of distribution shift or coarseness of time-frequency resolution for $h{=}N$ specifically. 

\subsection{Preserving performance in the original setting}
Unlike conventional fine-tuning approaches, which typically adapt a pretrained model to a new data domain without prioritizing original-task performance, our objective is to enable a single pretrained model to operate robustly across \textit{any} STFT window and hop length configuration $\mathcal{C}$. Thus, maintaining strong performance in the original setting is an important requirement for our adaptation strategy. To measure this, we evaluate performance of each model using the \textit{original} frontend, and report this result relative to the upper-bound: $\Delta \text{mAP}=\text{mAP}_{\text{trained}}-\text{mAP}_{\text{upper}}$. We show in Figure \ref{fig:robustness_and_preservation} (Right) that all of the FiLM-conditioned model variants are able to preserve performance in the original setting, with our best FiLM-Encoder version yielding an average mAP of $0.443$, only $-0.013$ from the upper-bound. The fine-tuning baselines suffer more when evaluated using original parameters, illustrating benefits of training a drop-in conditioning mechanism such as FiLM vs. traditional fine-tuning for frontend adaptation.

\begin{figure}
    \centering
    \includegraphics[trim={1cm 0.2cm 0.5cm 0cm},width=0.73\linewidth]{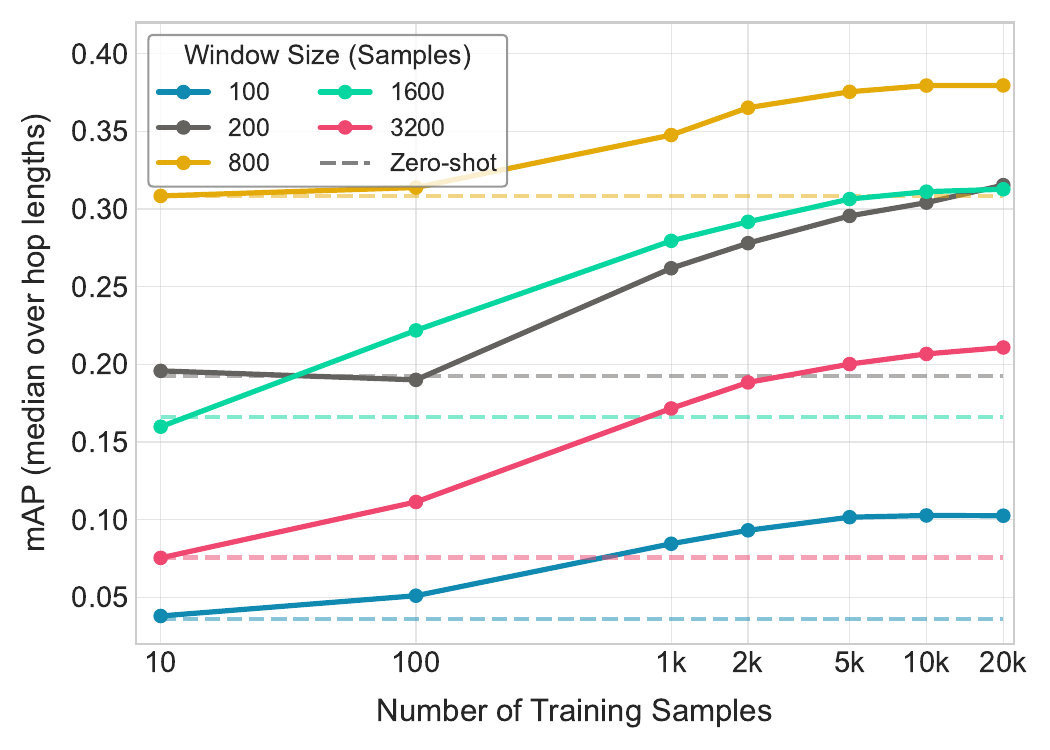}
    \vspace{-0.1cm}
    \caption{Few-shot frontend adaptation as a function of number of training samples (FiLM-Encoder variant). Full dataset training is denoted at $20$k samples.  
    }
    \label{fig:fewshot}
\end{figure}

\subsection{How much data do you need for FiLM adaptation?}
We evaluate the data efficiency of our best FiLM adaptation method (FiLM-Encoder) by varying the number of training samples used to learn the conditioning parameters. Figure \ref{fig:fewshot} shows that substantial improvements over the zero-shot AST baseline are achieved with as few as $1,000$ samples, with performance approaching that of the full dataset training ($20$k) using only $5$k training samples across window lengths. While the trend is fairly consistent across window lengths, for larger windows, e.g., $N{=}1600$, a relative improvement of $33.7\%$ ($+5.6$pp) over the zero-shot baseline is achievable with only 100 training samples.

\vspace{-0.1cm}
\section{Conclusion}
This study is an exploration of STFT frontend mismatch within a controlled setting, limited to the original AST model and pretraining dataset with variation in STFT window and hop length. Future work will expand to other audio transformer models, diverse datasets, and other parameters, e.g., number of Mel filters and sampling rate, as well as tasks beyond classification where sensitivity to time-frequency resolution is even more critical. Within this setting, we present a systematic analysis of STFT frontend mismatch in a pretrained Audio Spectrogram Transformer, and propose a lightweight FiLM-based conditioning strategy to mitigate its effects. Our findings show that audio classification performance degrades substantially as window and hop length deviate from pretraining settings, and that our approach recovers much of this degradation while preserving performance in the original configuration using as few as $1,000$ training samples.






\bibliographystyle{IEEEtran}
\bibliography{refs}


\end{document}